\documentstyle[epsf]{l-aa}
\newsavebox{\ggbox}
\newcommand{\displaymathbox}[1]%
{\savebox{\ggbox}{$\displaystyle #1$}\begin{displaymath}\fbox{\usebox{\ggbox}}%
\end{displaymath}}
\newcommand{\equationbox}[1]%
{\savebox{\ggbox}{$\displaystyle #1$}\begin{equation}\fbox{\usebox{\ggbox}}%
\end{equation}}

\newsavebox{\source}
\def\epigram{\@ifnextchar [{\@tempswatrue\@epigram}{\@tempswafalse\@epigram[]}}
\def\@epigram[#1]#2#3{\savebox{\source}%
{{\small\sf#2}\if@tempswa, {\small\it#1}\fi}\vspace{.2ex}%
\begin{flushright}\small\sl#3\vspace{.8ex}\usebox{\source}\end{flushright}}

\newcommand{\etal}[1]{et al.}

\newcommand{\firstauthor}[2]{#2 #1}
\newcommand{\otherauthor}[2]{#2 #1}

\newcommand{\article}[9]%
{\bibitem[#1]{#2}#3, #9, #5 \underline{#6}, #8}

\newcommand{\novolume}[8]%
{\bibitem[#1]{#2}#3, #8, #5 \underline{#6}, #7}

\def\edtext{eds.}
\newcommand{\proceedings}[9]%
{\bibitem[#1]{#2}#3, #9, #4.\, In: #6 (\edtext) #5. #8, #7}

\newcommand{\sproceedings}[9]%
{\bibitem[#1]{#2}#3, #9, #4.\, In: #6 (\edtext) #5. #8, #7}

\def\preptext{in preparation}
\newcommand{\prep}[4]%
{\bibitem[#1]{#2}#3, #4, \preptext}

\def\subtext{submitted}
\newcommand{\submitted}[6]%
{\bibitem[#1]{#2}#3, #6, #5 (\subtext)}

\def\accepttext{in press}
\newcommand{\inpress}[6]%
{\bibitem[#1]{#2}#3, #6, #5 (\accepttext)}

\newcommand{\book}[7]%
{\bibitem[#1]{#2}#3, #7, #4.\, #6, #5}

\newcommand{\doctor}[6]%
{\bibitem[#1]{#2}#3, #6, docotoral thesis, #5}

\newcommand{\laf}{Lyman-$\alpha$ forest}
\begin{document}
\thesaurus{
           02        % A&A Section
          (12.03.4;  % cosmology: theory
           11.17.1;  % quasars: absorption lines
           12.03.3;  % cosmology: observations
           12.12.1)  % large-scale structure of universe
}
\title{
       Evolution in the \laf
}
\author{
Thomas~Liebscher%\thanks{present address: }\thanks{email}
\and
Rainer~Kayser\thanks{\tt rkayser@hs.uni-hamburg.de} \and
Phillip~Helbig\thanks{\tt phelbig@hs.uni-hamburg.de}
}
\offprints{P.~Helbig}
\institute{Hamburger Sternwarte,
           Gojenbergsweg 112,
           D-21029 Hamburg, Germany
}
\date{received; accepted}
\maketitle
\markboth{Th.~Liebscher et al.: Evolution in the Lyman-$\alpha$ forest}{}
\begin{abstract}
We reanalyse the spectra used by \mbox{D.-E.} Liebscher et~al.\
(\cite{DLiebscherHP92b}) with the same goal -- determining the
cosmological parameters -- and basically the same assumptions but with a
different statistical method, which does not rely on binning the data.
Also, we correct for selection effects.  We basically confirm their
result, with somewhat larger but still very small (formal) errors.
However, all world models within the 99\% confidence region are ruled
out because a firm lower limit on $\Omega_{0}$ is significantly larger.
This directly demonstrates for the first time the existence of intrinsic
evolution in the Lyman-$\alpha$ forest. 
\keywords{cosmology: theory -- quasars: absorption lines -- 
          cosmology: observations -- large-scale structure of Universe}
\end{abstract}

%***********************************************************************
\section{Introduction}\label{intro}

A method of determining the cosmological parameters $\lambda_{0}$ and
$\Omega_{0}$ has recently been suggested by D.-E.~Liebscher
et~al.~(\cite{DLiebscherHP92b} -- hereafter LPH).  They use the redshift
distribution of \laf\ lines, with the assumption that the \laf\
absorbers have a constant comoving density.  This basically determines
the volume element as a function of redshift, which depends on the
cosmological model and is especially sensitive to a positive value of
the cosmological constant~$\lambda_{0}$.  (See, e.g.,
Feige~(\cite{BFeige92a}) for an explanation of this effect.)  Under this
assumption, which means no {\em intrinsic\/} evolution -- we henceforth
use `evolution' to mean intrinsic physical evolution and not in the more
general and somewhat misleading sense of `dependence on $z$' -- they
find an acceptable fit only in a small area of the
$\lambda_{0}$-$\Omega_{0}$-plane.  It is important to realise that, if
evolution is present in the absorbers, one would not necessarily expect
to find any (then necessarily wrong) cosmological model (not even an
extreme one perhaps excluded by other arguments) which would be
compatible with the observations without requiring evolution; i.~e.,
evolutionary effects and the influence of the cosmological model are not
necessarily degenerate, especially when, as is here the case, one has a
large data set extending to high redshifts.  Of course, the fact that
LPH find an acceptable fit does not prove the assumption of no
evolution, but it {\em does\/} mean that one {\em cannot\/} say that
`evidence of evolution is directly seen in the spectra'; the fact that
an acceptable fit is found demonstrates that a cosmological model exists
in which no evolution is required to explain the observed spectra. 

The paper by LPH was subjected to much criticism, but our investigation
shows that most of this is unfounded.  We have confirmed their results
with a better statistical method (and have reproduced their results with
their method as a check at the outset of our investigation); the result
is perfectly consistent with the assumptions and gives an acceptable
fit. 

This was admittedly somewhat surprising, since we have a number of
points of criticism of LPH.  First, the distribution of spectral lines
was not used directly, but rather the lines were counted in redshift
bins.  Binning introduces free parameters and throws away information
(cf., e.g., Press et~al.~(\cite{WPressTVF92a})) and should be avoided if
possible.  Also on the mathematical side, a fit to their $Z^{2}$ instead
of $Z$ was done.  In addition, since the smallest equivalent width
capable of being registered by {\em all\/} detectors used in the
inhomogeneous sample of \laf\ lines is 0.2~\AA, this implies, because
the lowest redshift in their sample is $z=0.003$, that the {\em
intrinsic\/} equivalent width $W_{\rm intr}$ to which the sample is
complete must also be $\approx 0.2\thinspace{\rm \AA}$. Of the 1297
lines used by LPH only 832 have a $W_{\rm intr}$ larger than this
minimum.  Furthermore, since the physical distance between two absorbers
is a function of their redshifts and the cosmological parameters
$\lambda_{0}$ and $\Omega_{0}$, their sample is only complete to
distances greater than that corresponding to the minimum observable
distance {\em in the given world model\/}.  To avoid biasing, we must
for each world model exclude a certain number of lines from the
statistical analysis.  (See the discussion of the correction for
`resolution bias' in Kayser~(\cite{RKayser95a}).)

%***********************************************************************
\section{Basic theory}

\subsection{Cosmology}

We make the `standard assumption' that the Universe can be described by
the Robertson-Walker metric.  With the assumption of no evolution, the
absorbers have a constant comoving density.  The proper distance (this
is the distance $D^{\rm P}$ in Kayser et~al.~(\cite{RKayserHS96a}),
hereafter KHS) in a given world model between two absorbers with
redshifts $z_{x}$ and $z_{y}$ is given by $D^{\rm P}(z_{y}) - D^{\rm
P}(z_{x})$.  This is simply 
\begin{equation}\label{dp-g}
D^{\rm P}_{xy} = \frac{c}{H_{0}}\int\limits_{z_{x}}^{z_{y}}
\frac{{\rm d}z}{\sqrt{Q(z)}}.
\end{equation}
where
\begin{equation}
\label{q-g}
Q(z) = \Omega_{0}(1 + z)^{3} - 
       (\Omega_{0} + \lambda_{0} -1)(1 + z)^{2} +
       \lambda_{0} .
\end{equation}
The idea is to find the cosmological parameters for which $D^{\rm
P}_{xy}$ corresponds most closely to what is observed in the spectra; if
there is no evolution, then this must be the correct world model, if the
fit is acceptable.  We use the distance between adjacent
absorbers~$\Delta z$ as the observational quantity.  This is of course
equivalent to the number of lines per redshift interval except for the
loss of precision due to binning in the latter case.

\subsection{Statistics}

We use a method similar to that described in Kayser~(\cite{RKayser95a}):
We calculate the distribution of distances between pairs of adjacent
absorbers, divide the sample into $N$ subsamples based on redshift (with
the same number of objects in each subsample) and compare \mbox{$m =N/2$} 
independent pairs of samples with $m$ Kolmogorov-Smirnov
(\mbox{K-S}) tests (instead of just two samples and one test as in
Kayser~(\cite{RKayser95a})\footnote{It is obvious that using just two
subsamples will not allow one to reject the null hypothesis that the
distribution is independent of redshift if there is a symmetry with
respect to the redshift separating the subsamples, even though the 
distribution is not independent of redshift.}).  Since the null
hypothesis is that the distribution is independent of redshift, the
choice of independent pairs is arbitrary. 

\begin{figure}[t]
\epsffile{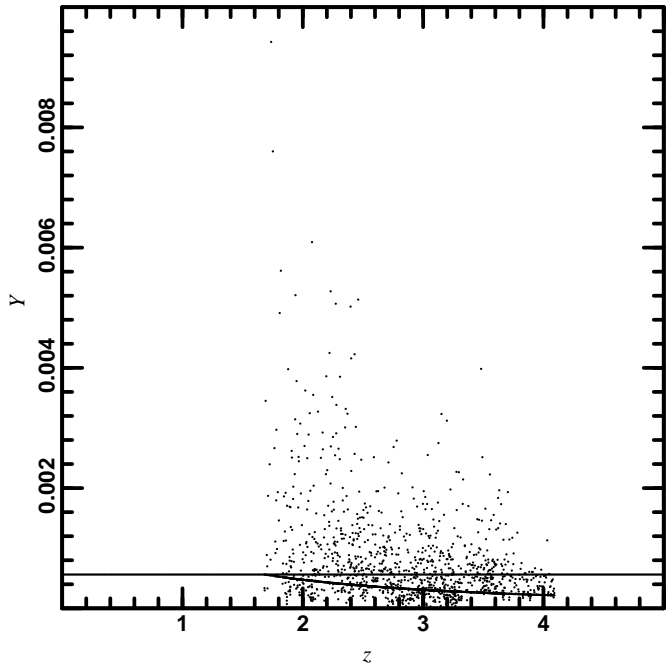}
\caption[]{Demonstration of the statistical method.  Plotted is the
comoving distance Y in units of $c/H_{0}$ between pairs of adjacent
absorbers as a function of redshift.  Also shown is the minimum
observable distance (lower curve) in the given world model in the
interval where data points exist; the maximum of this curve (in the
interval where data are used) determines the (straight horizontal) line
below which the data cannot be used if one requires a complete sample.
Here, $\lambda_{0} = 0$ and $\Omega_{0} = 1$.  Pairs of subsamples in
different redshift intervals are compared with  Kolmogorov-Smirnov tests
and a probability is calculated that the different subsamples could have
been drawn from the same parent population, thus giving the probability
that the distribution is independent of redshift} 
\label{ks-fig}
\end{figure}

What is the probability $P_{\rm min}$ that the minimum of all $m$
\mbox{K-S} probabilities $P_{i}$ is less than a given $P_{0}$?  Since,
if $\min(P_{i}) > P_{0}$ {\em all\/} values $P_{i}$ must be larger than
$P_{0}$ we have 
\begin{equation}
P(\min(P_{i}) > P_{0}) = (1 - P_{0})^{m}
\end{equation}
and thus
\begin{equation}
P_{\rm min} = P(\min(P_{i}) < P_{0}) = 1 - (1 - P_{0})^{m}
\quad .
\end{equation}
Solving for $P_{0}$ we obtain
\begin{equation}
P_{0} = 1 - \sqrt[m]{1 - P_{\rm min}}
\end{equation}
We can reject the null hypotheses that this distribution is independent
of redshift at the $100(1 - P_{\rm min})\%$ confidence level in the
cases where $\min(P_{i}) < P_{0}$. Here, we conservatively considered a
probability $P_{\rm min}$ of $1\%$ or larger to be compatible with the
observations and used three pairs of samples, thus ruling out all world
models with a value of $\min(P_{i})$ less than $P_{0} =
1-\sqrt[3]{0.99}\approx 0.0033$ at the $99\%$ confidence level. Of
course, as mentioned above, for each world model the actual sample is
different in order that it be complete down to a certain distance
between pairs of adjacent absorbers. (See Fig.~\ref{ks-fig}.) 
%%%%%%%%%%%%%%%%%%%%%%%%%%%%%%%%%%%%%%%%%%%%%%%%%%%%%%%%%%%%%%%%%%%%%%%%
%%%%%%%%%%%%%%%%%%%%%%%%%%%%%%%%%%%%%%%%%%%%%%%%%%%%%%%%%%%%%%%%%%%%%%%%
%%%%%%%%%%%%%%%%%%%%%%%%%%%%%%%%%%%%%%%%%%%%%%%%%%%%%%%%%%%%%%%%%%%%%%%%

%***********************************************************************
\section{The sample}

For purposes of comparison, we used the same sample of spectral lines as
in LPH, except that we omitted 7 lines with $z<1.65$ (taken from Bahcall
et al. (\cite{JBahcallJSetal91a})).  This means that the sample is now
complete to 
\begin{displaymath}
W_{\rm intr}=0.2/(1 + 1.65) \approx 0.0755\thinspace{\rm
\AA}
\quad ,
\end{displaymath}
which gives us a sample consisting of 1108 of the 1297 LPH lines;
including these 7 low-redshift lines would decrease the complete sample,
as explained in Sect.~\ref{intro}, to 832 lines.  Although the
low-redshift lines have a significant impact on the $\chi^{2}$ method
used by LPH, leaving out this few lines makes practically no difference
in the {\em statistic\/} we use, so we choose to leave them out in order
to be able to work with a larger complete sample.  Since we wanted to
use the actual redshifts of the lines and not simply the number of lines
per redshift interval, the wavelengths of the lines were determined from
scans of the literature spectra and the redshifts (and $W_{\rm intr}$)
were then calculated. See Table~\ref{table} for more details. 
%%%%%%%%%%%%%%%%%%%%%%%%%%%%%%%%%%%%%%%%%%%%%%%%%%%%%%%%%%%%%%%%%%%%%%%%
%%%%%%%%%%%%%%%%%%%%%%%%%%%%%%%%%%%%%%%%%%%%%%%%%%%%%%%%%%%%%%%%%%%%%%%%
\begin{table}[t]
\caption[]{Spectra used.  The redshift $z$ refers to the midpoint of the 
wavelength interval.  References are Atwood et
al.\ (\cite{BAtwoodBC85a}), Pettini et al.\ (\cite{MPettiniHSM90a}), Sargent et
al.\ (\cite{WSargentBS88a}), Steidel\ (\cite{CSteidel90a}), Carswell et
al.\ (\cite{RCarswellLPW91a}) and Rauch et al.\ (\cite{MRauchCCetal92a})} 
\label{table}
\begin{flushleft}
%\begin{tabular}{|r|l|l|c|l|l|}
\begin{tabular}{|r|l|c|c|l|l|}
\hline
&&&&&\\
\# & 
QSO & 
z
&
$\lambda$ [${\rm \AA}$] & 
$\frac{\Delta\lambda}{\lambda}$ & 
reference  \\
&&&&&\\
\hline
&&&&&\\
%1 & 3C273 & 0.08 & 1216-1408 & & Bahcall \\
1 & 0420-388 & 2.25 & 3850-4050 & 0.5 & Atwood \\
2 & 0420-388 & 2.46 & 4100-4300 & 0.5 & Atwood \\
3 & 0420-388 & 2.46 & 4450-4650 & 0.5 & Atwood \\
4 & 0420-388 & 2.99 & 4750-4950 & 0.5 & Atwood \\
5 & 2206-199N & 2.25 & 3860-4052 & 0.09 & Pettini \\
6 & 2206-199N & 2.43 & 4070-4282 & 0.09 & Pettini \\
7 & 0114-089 & 3.11 & 4900-5100 & 1.5 & Sargent \\
8 & 0913+072 & 2.45 & 3900-4500 & 1.5 & Sargent \\
9 & 1159+124 & 3.11 & 4600-5400 & 1.5 & Sargent \\
10 & 1247+267 & 1.84 & 3260-3660 & 0.8 & Sargent \\
11 & 1511+091 & 2.70 & 4300-4700 & 1.5 & Sargent \\
12 & 1623+269 & 2.29 & 3700-4300 & 0.8 & Sargent \\
13 & 2126-158 & 3.11 & 4800-5200 & 1.5 & Sargent \\
14 & 0142-100 & 2.45 & 3900-4500 & 0.8 & Sargent \\
15 & 0237-233 & 2.04 & 3500-3900 & 0.8 & Sargent \\
16 & 0424-131 & 2.04 & 3600-3800 & 0.8 & Sargent \\
17 & 1017+280 & 1.80 & 3300-3500 & 0.8 & Sargent \\
18 & 2000-330 & 3.44 & 5100-5700 & 1.35 & Steidel \\
19 & 0055-209 & 3.52 & 5400-5600 & 1.35 & Steidel \\
20 & 0000-263 & 3.77 & 5400-6200 & 1.35 & Steidel \\
21 & 1208+101 & 3.60 & 5400-5800 & 1.51 & Steidel \\
22 & 1100-264 & 1.91 & 3440-3640 & 0.11 & Carswell \\
23 & 1100-264 & 2.05 & 3640-3780 & 0.11 & Carswell \\
24 & 0014+813 & 2.74 & 4500-4600 & 0.36 & Rauch \\
25 & 0014+813 & 2.87 & 4600-4800 & 0.36 & Rauch \\
26 & 0014+813 & 3.03 & 4800-5000 & 0.36 & Rauch \\
27 & 0014+813 & 3.15 & 5000-5100 & 0.36 & Rauch \\
28 & 0014+813 & 3.28 & 5100-5300 & 0.36 & Rauch \\
&&&&&\\
\hline
\end{tabular}
\end{flushleft}
\end{table}
%%%%%%%%%%%%%%%%%%%%%%%%%%%%%%%%%%%%%%%%%%%%%%%%%%%%%%%%%%%%%%%%%%%%%%%%
Our measured wavelengths of all the lines are available from the
following URLs: 
\begin{quote}
\scriptsize \tt
 http://www.hs.uni-hamburg.de/english/persons/helbig/\\
        Research/Publications/Info/lyman-alpha.html
\end{quote}
\begin{quote}
\scriptsize \tt
 ftp://ftp.uni-hamburg.de/pub/misc/astronomy/\\
       lyman-alpha.tar.gz
\end{quote}

%***********************************************************************
\section{Calculations, results and discussion}

\begin{figure}[t]
\epsffile{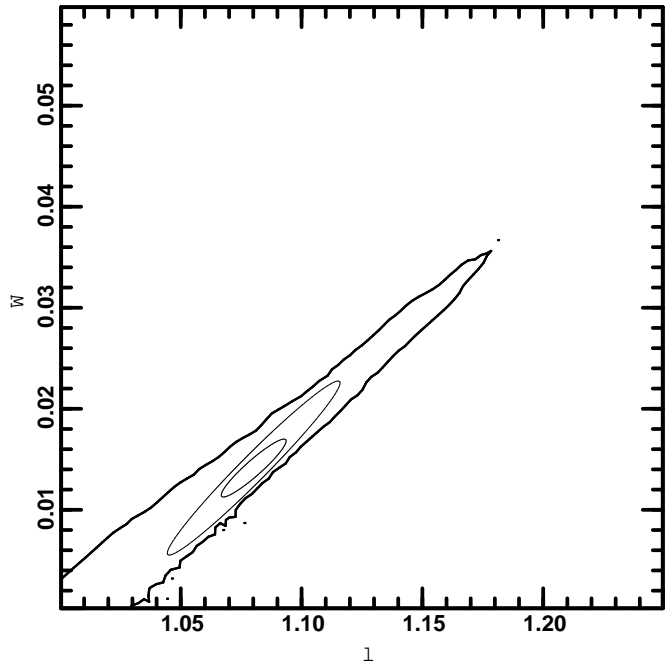}
\caption[]{The probability as a function of the cosmological parameters.
The (thick) contour line is at 1\%. Only world models within this area
are compatible with the data.  Note the extremely small area of the
$\lambda_{0}$-$\Omega_{0}$-plane in this plot -- outside this area, {\em
no\/} areas with an acceptable probability exist.  The lower left corner
of the plot corresponds to the de~Sitter model.\footnote{Most of the
space to the lower right of the allowed area corresponds to world models
in which the maximum possible redshift is less than the largest
Lyman-$\alpha$ redshift in the sample.  These are the so-called bounce
models, which have no big bang and, for realistic values of
$\Omega_{0}$, are excluded by the mere existence of objects with a large
cosmological redshift (see Stabell \&
Refsdal~(\cite{RStabellSRefsdal66a}) or KHS).  For low $\Omega_{0}$
values there are some bounce models at the lower part of the allowed
area which are not excluded by the redshift of the lines used in this
analysis.}  For comparison, the LPH 1~$\sigma$ and 3~$\sigma$ error
ellipsoids are included} 
\label{result-fig}
\end{figure}

First, for a general overview, we calculated the probability for
cosmological models in the area given by \mbox{$-10 < \lambda_{0} < 7$}
and \mbox{$0 < \Omega_{0} < 7$}, with a resolution of 0.06 \ ($\approx$
33~000 different models).  Almost the entire area of this plane is
rejected.  Then, in a series of steps, the area containing a probability
$P_{\rm min} > 1\%$ was examined in progressively higher resolution. 
Only a small region, shown in Fig.~\ref{result-fig}, was found to be
compatible with the observational data. 
%%%%%%%%%%%%%%%%%%%%%%%%%%%%%%%%%%%%%%%%%%%%%%%%%%%%%%%%%%%%%%%%%%%%%%%%
%%%%%%%%%%%%%%%%%%%%%%%%%%%%%%%%%%%%%%%%%%%%%%%%%%%%%%%%%%%%%%%%%%%%%%%%
\footnotetext{Most of the space to the lower right of the allowed area
corresponds to world models in which the maximum possible redshift is
less than the largest Lyman-$\alpha$ redshift in the sample. Most of
these are the so-called bounce models, which have no big bang and, for
realistic values of $\Omega_{0}$, are excluded by the mere existence of
objects with a large cosmological redshift (see Stabell \&
Refsdal~(\cite{RStabellSRefsdal66a}) or KHS).  For low $\Omega_{0}$
values there are some bounce models at the lower part of the allowed
area which are not excluded by the redshift of the lines used in this
analysis.} 
%%%%%%%%%%%%%%%%%%%%%%%%%%%%%%%%%%%%%%%%%%%%%%%%%%%%%%%%%%%%%%%%%%%%%%%%

Thus, we find in agreement with LPH that -- with the assumption of no
evolution -- cosmological models exist which provide an acceptable fit
to the data.  The fact that the area in the
$\lambda_{0}$-$\Omega_{0}$-plane in which this is the case is so small
means that, if the method is correct, it offers by far the best known
method of determining the cosmological parameters. The fact that the
allowed area is somewhat outside the (present) canonical range of
cosmological models is probably one reason for some of the criticism of
the work of LPH.  Mostly, this has concentrated on the fact that this
cosmological model is dominated by the cosmological constant.  We hasten
to point out that it has not yet been shown that there is a way of
excluding this large a value of $\lambda_{0}$ in spatially closed
($k=+1$) world models.\footnote{All cosmological models in
Fig.~\ref{result-fig} have $k=+1$ and $\lambda_{0}>1$ and thus are
spatially closed and will expand forever. See KHS.} It is generally
recognised that gravitational lensing statistics can provide the
tightest constraints on a positive cosmological constant. However, at
present such constraints have only been derived for {\em flat\/} ($k=0$)
models.  (See, e.g., Kochanek~(\cite{CKochanek96a}) and references
therein.)  Basically, large $\lambda_{0}$ values predict too many
lenses, primarily because of the relatively large volume element ${\rm
d}V/{\rm d}z$ in these models.  For $k=+1$ models, this effect is not as
pronounced, since ${\rm d}V/{\rm d}z$ is generally smaller in a $k=+1$
model than in a flat model with the same value of $\lambda_{0}$.  Thus,
a given upper limit on $\lambda_{0}$ for a flat model corresponds to a
{\em weaker\/} limit for a $k=+1$ model. Nevertheless, it is conceivable
that lensing statistics could rule out the LPH\footnote{In the
following, the term `LPH model' should be understood to mean the range
of cosmological models within the errors of the LPH best fit model {\em
as well as our results\/} since, for the purposes of the following
discussion, the cosmological models are identical.} model and other $k =
+1$ models with a large cosmological constant; this is currently under
investigation. 

LPH point out `advantages' of their model, such as the fact that a
relatively old universe and a relatively large Hubble constant can both
be accommodated.  In fact, better values for the world age and $H_{0}$
will probably rule out some standard cosmological models in the near
future.  The point is that such indirect tests don't rule out the LPH
model; if anything, they point in the direction of this or a similar
cosmological model. 

There is not necessarily any non-baryonic matter in the LPH model, since
their value of $\Omega_{0}$, coupled with plausible values for $H_{0}$,
gives a density comparable to the baryonic density predicted by
nucleosynthesis.\footnote{For some `more standard' low-density standard
cosmological models, it is also just barely possible to have no
non-baryonic matter, but only for an extremely low value of $H_{0}$ and
a baryonic density at the upper limit.} (See, e.g., Krauss
(\cite{LKrauss95a}) and references therein.)  This could also be seen as
a point in favour of this cosmological model. 

However, the $\approx 3\sigma$ upper limit for the {\em total density\/}
$\Omega_{0}$ ($\approx 0.035$) is significantly less than the {\em
extremely\/} conservative lower limit ($\approx 0.05$) advocated by
Peebles (\cite{PPeebles93a}, Chapter~20).  Since this conservative lower
limit on $\Omega_{0}$, based on well-understood physics, is considerably
less than lower limits advocated by other researchers, this means that
the LPH model is definitively ruled out and that it is not possible to
determine the cosmological parameters by this method.

%***********************************************************************
\section{Evolution}

The conclusion in the last paragraph means that there {\em must\/} be
evolution in the \laf, because without evolution, the only cosmological
model compatible with the observations is definitively ruled out on
other grounds.  {\em This conclusion does not assume any a priori values
for the cosmological parameters.} 

\begin{figure}[t]
\epsffile{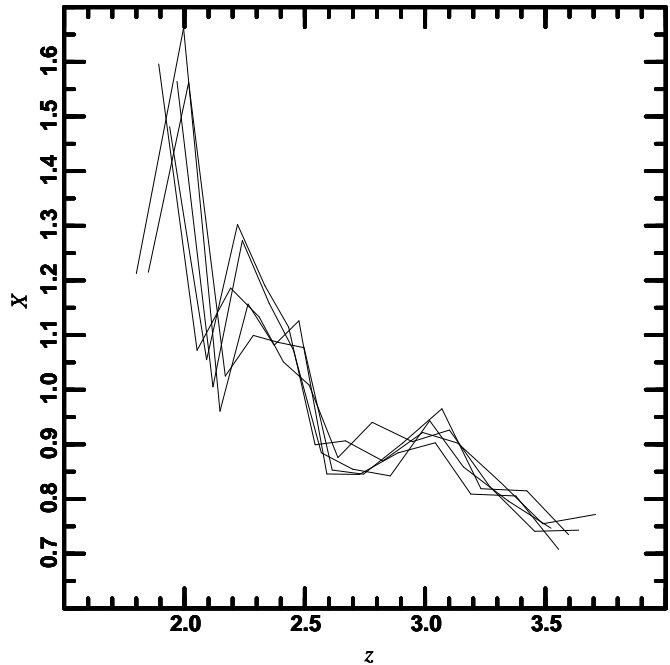}
\caption[]{Evolution in the \laf\ for the Einstein-de~Sitter model
($\lambda_{0}=0$, $\Omega_0=1$).  Higher values mean greater distances
between absorbers and thus {\em fewer\/} absorbers per comoving volume. 
(Values are normalised with respect to the mean.) Different curves
correspond to different binning schemes} 
\label{evolution1-fig} 
\end{figure}

For a given cosmological model, the comoving distances between adjacent
\laf\ absorbers can be calculated, as shown above.  A plot of the
average distance per redshift bin should be almost flat if the
cosmological model used for the calculation is correct and if there is
no evolution. Previously, we assumed no evolution to so determine the
cosmological model.  However, since we now know that there must be
evolution in the \laf, we now, for a few representative cosmological
models, calculate what evolution is necessary to explain the
observations.  Several different binning schemes were used to isolate
real effects from noise. (See Fig.~\ref{evolution1-fig}). 
%%%%%%%%%%%%%%%%%%%%%%%%%%%%%%%%%%%%%%%%%%%%%%%%%%%%%%%%%%%%%%%%%%%%%%%%
%%%%%%%%%%%%%%%%%%%%%%%%%%%%%%%%%%%%%%%%%%%%%%%%%%%%%%%%%%%%%%%%%%%%%%%%
%%%%%%%%%%%%%%%%%%%%%%%%%%%%%%%%%%%%%%%%%%%%%%%%%%%%%%%%%%%%%%%%%%%%%%%%
%%%%%%%%%%%%%%%%%%%%%%%%%%%%%%%%%%%%%%%%%%%%%%%%%%%%%%%%%%%%%%%%%%%%%%%%
%%%%%%%%%%%%%%%%%%%%%%%%%%%%%%%%%%%%%%%%%%%%%%%%%%%%%%%%%%%%%%%%%%%%%%%%
\begin{figure}[t]
\epsffile{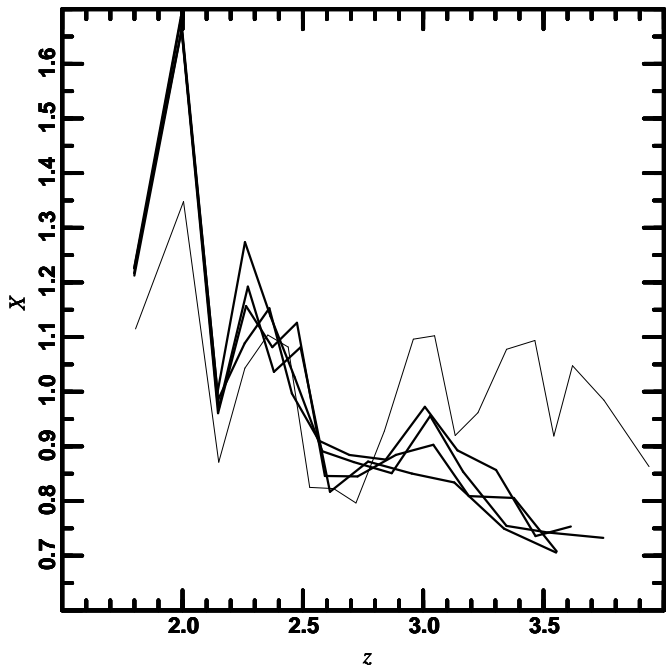}
\caption[]{The same as Fig.~\ref{evolution1-fig} but with one binning
scheme and different curves correspond to different cosmological models.
The thick curves correspond to ($\lambda_{0}$,$\Omega_{0}$) values of
(0.0,1.0) (0.0,0.3), (0.65,0.35) and (2.0,4.0).  The curves are
practically the same within the binning errors
(cf.~Fig~\ref{evolution2-fig}).  The thin curve corresponds to the
best-fit model of HPL, ($\lambda_{0}$,$\Omega_{0}$) = (1.080,0.014), for
which of course the calculated evolution curve is almost flat within the
binning errors.} 
\label{evolution2-fig} 
\end{figure}
%%%%%%%%%%%%%%%%%%%%%%%%%%%%%%%%%%%%%%%%%%%%%%%%%%%%%%%%%%%%%%%%%%%%%%%%
The resulting calculated evolution is relatively independent of the
world model (except near the LPH model, where of course the calculated
evolution is almost flat).  This means that Fig.~\ref{evolution1-fig} is
approximately correct for almost any world model
(cf.~Fig.~\ref{evolution2-fig}).  This is the first calculation of the
evolution of the \laf\ based on observational data which makes no
assumptions (other than basic cosmological ones).  The general trend is
an {\em increase\/} in the comoving density of \laf\ absorbers with $z$
and thus a {\em decrease\/} with time. 

The form of the curves in Fig.~\ref{evolution1-fig} make it possible to
understand why the LPH model is compatible with the observations without
requiring any evolution.  The basic feature in the curves in
Fig.~\ref{evolution1-fig} is the fact that the comoving distance between
adjacent absorbers decreases with increasing redshift.  If there were no
evolution, then the comoving distance should be independent of redshift.
Thus, a decrease in the comoving distance between adjacent absorbers
with increasing redshift can be compensated in a world model in which
the comoving distance for a fixed $\Delta z$ is comparatively larger at
larger redshifts.  This is only the case in a small part of the
$\lambda_{0}$-$\Omega_{0}$-plane. 

For a given $\Omega_0$ value there is a value of $\lambda_{0}$ above
which no big bang is possible; rather, the universe contracts from
$\infty$ to a finite $R_{\rm min}$ before expanding again.  Near this
curve separating world models with and without a big bang, which for
brevity we refer to in the following as the $A_{2}$ curve\footnote{This
is {\tt WMTYPE 17} as given by the {\tt INICOS} routine mentioned in
KHS.} (following Stabell \& Refsdal~(\cite{RStabellSRefsdal66a})) are
the so-called plateau models, where the universe has a quasi-static
phase.  (See, e.g. Feige~(\cite{BFeige92a}) for a discussion of this
phase.)  Near the redshift of the plateau, the comoving distance for a
fixed $\Delta z$ can become very large.  The redshift of the plateau
depends sensitively on the value of the cosmological parameters, ranging
from $\infty$ near the de~Sitter model ($\lambda_{0}=1$, $\Omega_{0}=0$)
to values $<1$ for relatively small values of $\Omega_{0}$ and
$\lambda_{0}$. Figure~\ref{evolution1-fig} shows the comoving distances
between adjacent absorbers becoming smaller with increasing redshift,
here assumed to be due to intrinsic evolution. It is easy to see that a
cosmological model -- such as the ones near the LPH model -- in which
the distance is relatively larger at larger redshift will counteract
this effect, producing a flatter distribution which, in such a
cosmological model, would imply a lack of intrinsic evolution.  Since
the redshift of the plateau and hence of the large volume element can be
adjusted by moving along the $A_{2}$ curve, and its width by moving
perpendicular to it, it is easy to find a cosmological model which, with
no evolution, will be compatible with observational data from a `normal'
cosmological model in which the comoving distance between absorbers for
a fixed $\Delta z$ decreases with increasing redshift due to
evolutionary effects.  Due to the sensitivity mentioned above, this will
only be possible in a very small area of the parameter space. Since the
{\em width\/} of the plateau is even more sensitive to the distance from
the $A_{2}$ curve than is its redshift to the position along the curve,
the allowed area in Fig.~\ref{result-fig} has an elongated shape.

%***********************************************************************
\section{Summary and conclusions}

We confirm the result of D.-E.~Liebscher
et~al.~(\cite{DLiebscherHP92b}): with the assumption of no evolution,
observations of the \laf\ allow only a small range of (spatially closed)
cosmological models with a positive cosmological constant and a low
value of $\Omega_{0}$.  This is despite the fact that we avoided some of
the pitfalls of their method such as binning and selection effects.  The
former is relatively harmless in this case because of the large number
of data points; the fact that the latter isn't important is probably
just coincidence. 

However, the fact that all cosmological models within the errors of the
`best fit' model are ruled out because a firm lower limit on
$\Omega_{0}$ is significantly larger means that there {\em must\/} be
intrinsic evolution in the \laf.  Calculating the necessary evolution
for a variety of cosmological models gives similar results, so that one
can make some qualitative conclusions about the \laf\ with almost no
knowledge about the cosmological model: the comoving distance between
adjacent absorbers decreases with redshift (and thus increases with
time).  This is a result which is very robust in that, in the area of
parameter space which is not excluded on other grounds, it is almost
independent of the cosmological model.

%***********************************************************************

\begin{acknowledgements}
It is a pleasure to thank S.~Refsdal for helpful discussions.  
\end{acknowledgements}

%***********************************************************************

%***********************************************************************

\end{document}